\documentclass[a4paper,11pt]{article}
\usepackage{pos}

\usepackage{multirow}
\usepackage{tikz}
\usepackage{wrapfig}

\DeclareMathOperator{\Tr}{\mathrm{Tr}}

\DeclareMathOperator{\SU}{\mathrm{SU}}

\newcommand{\real}{{\rm Re\,}}

\title{Intrinsic width of the flux tube in 2+1 dimensional Yang-Mills theories}

\author*[a]{Lorenzo Verzichelli}
\author[a]{Michele Caselle}
\author[a, b]{Elia Cellini}
\author[a]{Alessandro Nada}
\author[a]{Dario Panfalone}

\affiliation[a]{Universit\`a degli studi di Torino \& INFN -- sezione di Torino ,\\
  Via Pietro Giuria 1, Torino, Italy}

\affiliation[b]{Higgs Centre for Theoretical Physics \& School of Physics and Astronomy, The University of Edinburgh \\
  Edinburgh EH9 3FD, United Kingdom}

\emailAdd{lorenzo.verzichelli@unito.it}

\abstract{We present our updated results on the intrinsic width of the profile of the flux tube in (2+1)-dimensional Yang-Mills theory with $\SU(2)$ gauge group. We identify the intrinsic width as the characteristic length scale of the exponentially decaying tails of the profile of the flux tube. Inspecting a broad range of temperature, we check that this length does not depend on the length of the flux tube. Our estimations of the intrinsic width show a constant value at low temperature and a growing trend approaching the deconfinement temperature that can be understood from the universality class of the phase transition via the Svetitsky-Yaffe mapping.}

\FullConference{The 42nd International Symposium on Lattice Field Theory (LATTICE2025)\\
2-8 November 2025\\
Tata Institute of Fundamental Research, Mumbai, India\\}

\begin{document}
\maketitle

\section{Introduction}
An important feature of confining gauge theories is the formation of a flux tube between two colored sources, where the chromo-electric and chromo-magnetic field are confined. The dynamics of the flux tube at low energy can be described in terms of an effective string theory, which neglects its internal degrees of freedom and only describes its position in space. In this picture, the flux tube has a finite width only due to the fluctuations of the effective string. Such fluctuations are expected to lead to a Gaussian profile and this expectation is matched by recent numerical computations with advanced deep-learning algorithms~\cite{Caselle:2023mvh, Caselle:2024ent} both when the effective string action is the Nambu-Got\=o one, and when also including corrections proportional to powers of the extrinsic curvature.

However, it is known that the profile of the flux tube in confining gauge theories exhibits deviations from a Gaussian (see Ref.~\cite{Caselle:2016mqu} for numerical evidences), due to the intrinsic degrees of freedom of the theories. In the case of $\mathrm{U}(1)$ in (2+1) dimensions, the confinement is analytically understood~\cite{Polyakov:1975rs} and the deviations from a Gaussian profile can be explained in terms of interactions between the flux tube with the lightest excitations of the bulk theory~\cite{Aharony:2024ctf}. In that scenario, the profile is, at leading order in the gauge coupling, the convolution between the Gaussian fluctuations of the flux tube (described by the effective string theory) and an intrinsic shape that can be related to a solitonic classical solution. The latter exhibits exponentially decaying tails, which decrease more slowly than the Gaussian contribution; consequently, the convolution inherits an overall exponential fall-off at large distances. The characteristic length in those tails, that we will denote by $\lambda$, is only due to the intrinsic part of the profile and thus earns the name \textit{intrinsic width}. 

In the non-abelian case, we lack a complete analytical understanding of the confining mechanism and thus a description of the flux tube. However, there are hints pointing to the emergence of features analogous to those we just described. For example, the holographic duality between confining gauge theories at large number of colors $N_c$ and a gravity theory in AdS predicts a profile featuring exponentially decaying tails with an intrinsic characteristic length equal to the inverse of the mass-gap of the theory~\cite{Danielsson:1998wt, Canneti:2025afi}. Also the dual superconductor picture of confinement~\cite{Mandelstam:1974pi, tHooft:1981bkw} leads to exponentially decaying tails in the profile, with a characteristic length that would coincide with the London penetration depth of the dual superconductor. However there is no prediction on what such a length should be.

In our works \cite{Verzichelli:2025cqc, Caselle:2026coc} we shed some light on the non-abelian case far from the large $N_c$ limit, specifically for $\SU(2)$ in (2+1) dimensions. Our aim is to produce numerical data precise enough to be able to support or disprove predictions for the profile of the flux tube that come from a specific model for confinement, such as the dual superconductor or the holographic duality. All the numerical results we produced can be found in Ref.~\cite{Caselle:2026coc}. In particular, here we present some results for the profile in a broad range of temperatures. We show how the low temperature data can be fitted by a model originated from the dual superconductor model, leading however to some inconsistencies with the broader dual superconductivity picture. We also give an estimation of the intrinsic width at low temperature and compare it with the results obtained inspecting the entanglement entropy of the flux tube~\cite{Amorosso:2024glf, Amorosso:2024leg, Amorosso:2026mdo}.

Approaching the deconfinement temperature $T_c$, we show how the universality class of the phase transition allows for robust predictions on the profile of the flux tube via the Svetitsky-Yaffe mapping~\cite{Svetitsky:1982gs}. We test these predictions numerically, finding good agreement even at $T = 0.68 \, T_c$. 

\section{Lattice Setup}
We consider the (2+1)-dimensional $\SU(2)$ Yang-Mills theory regularized on a cubic lattice of spacing $a$ with $N_s$ sites in each of the two space directions and $N_t$ in the time one. For the discretized gauge action we choose the Wilson formulation:
\begin{equation}
    S[U] = \beta \sum_{x, \, \mu < \nu} \left( 1 - \real \Pi_{\mu \nu}(x) \right),
\end{equation}
where $\beta$ is related to the inverse bare gauge coupling as $\beta = 4 / (a g^2)$, while $\Pi_{\mu\nu}$ is the plaquette operator. Note that we absorbed a factor $1 / N_c = 1 / 2$ in the definition of the plaquette: 
\begin{equation}
    \Pi_{\mu \nu} = \frac{1}{2} \, \Tr [U_\mu(x) U_\nu(x+\hat{\mu}) {U_\mu}^\dagger(x+\hat{\nu}) {U_\nu}^\dagger(x)].
\end{equation}

Similarly, we define the Polyakov loop as
\begin{equation}
    P(\vec{x}) = \frac{1}{2} \Tr \left[ \prod_{t = 1}^{N_t} U_0(\vec{x}, t) \right].
\end{equation}

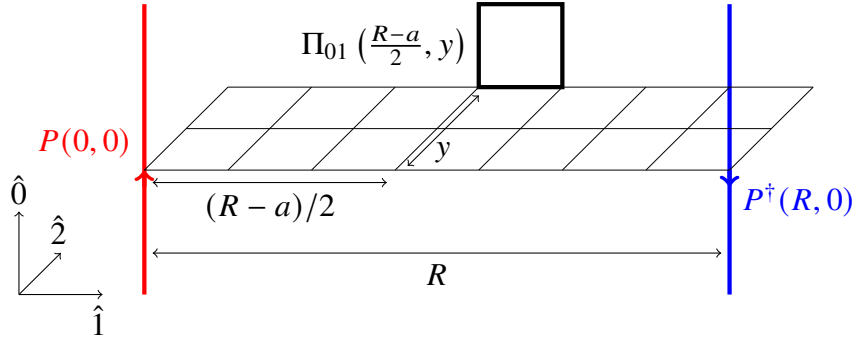
\begin{figure}
    \centering
    \begin{tikzpicture}[scale=1.1, transform shape]

    \draw [ -> ] (-1, 0) -- (-1, 1) node [pos=0.95, anchor=south] {$\hat 0$};
    \draw [ -> ] (-1, 0) -- (0, 0) node [pos=0.95, anchor=north] {$\hat 1$};
    \draw [ -> ] (-1, 0) -- (-0.5, 0.5) node [pos=0.95, anchor=south] {$\hat 2$};

    \draw [red, ultra thick, ->] (0.5, 0) -- (0.5, 1.5) node [anchor=south east] {$P(0, 0)$};
    \draw [red, ultra thick] (0.5, 1.3) -- (0.5, 3.5);

    \draw [blue, ultra thick] (7.5, 0) -- (7.5, 1.5) node [anchor=north west] {$P^\dagger(R, 0)$};
    \draw [blue, ultra thick, ->] (7.5, 3.5) -- (7.5, 1.3);

    \draw [ultra thin] (0.5, 1.5) -- (7.5, 1.5);
    \draw [ultra thin] (1.0, 2.0) -- (8.0, 2.0);
    \draw [ultra thin] (1.5, 2.5) -- (8.5, 2.5);

    \draw [ultra thin] (0.5, 1.5) -- (1.5, 2.5);
    \draw [ultra thin] (1.5, 1.5) -- (2.5, 2.5);
    \draw [ultra thin] (2.5, 1.5) -- (3.5, 2.5);
    \draw [ultra thin] (3.5, 1.5) -- (4.5, 2.5);
    \draw [ultra thin] (4.5, 1.5) -- (5.5, 2.5);
    \draw [ultra thin] (5.5, 1.5) -- (6.5, 2.5);
    \draw [ultra thin] (6.5, 1.5) -- (7.5, 2.5);
    \draw [ultra thin] (7.5, 1.5) -- (8.5, 2.5);

    \draw [ultra thick] (4.5, 2.5) -- (5.5, 2.5) -- (5.5, 3.5) -- (4.5, 3.5) -- cycle;
    \node at (4.5, 3) [anchor=east] {$\Pi_{01} \left( \frac{R - a}{2}, y \right)$};

    \draw [ <-> ] (0.6, 0.5) -- (7.4, 0.5) node [pos=0.5, anchor=north] {$R$};
    \draw [ <-> ] (0.6, 1.35) -- (3.4, 1.35) node [pos=0.5, anchor=north] {$(R - a) / 2$};
    \draw [ <-> ] (3.65, 1.55) -- (4.5, 2.4) node [pos=0.5, anchor=north] {$y$};

\end{tikzpicture}
    \vspace{-1.2 cm}
    \caption{Schematic representation of the three point function $F_{01}(R, y)$.}
    \label{fig:three-pt_schem}
\end{figure}

In order to probe the flux-tube~\cite{Fukugita:1983du, DiGiacomo:1990hc}, we consider the following (disconnected) three-point function:
\begin{equation}
    F_{\mu \nu}(R, y) = \left< \frac{1}{{N_s}^2} \, \sum_{\vec{x}} P \! \left(\vec{x}\right) \, \Pi_{\mu\nu} \! \left(\vec{x} + \vec{l} \right) \, P^\dagger \! \left(\vec{x} + R \hat 1\right) \right>,
\label{eq:three_pts}
\end{equation}
where $R$ is an odd multiple of the lattice spacing $a$ and $\vec{l}$ has (integer in units of $a$) components $(R - a) / 2$ along the direction $\hat 1$ and $y$ along $\hat 2$ (see Fig.~\ref{fig:three-pt_schem} for a schematic depiction). 

In order to define an observable whose continuum limit is the trace of the square modulus of the field strength, we normalize the three point function by the usual two point function
\begin{equation}
    G(R) = \left< \frac{1}{{N_s}^2} \, \sum_{\vec{x}} P \left(\vec{x}\right) \, P^\dagger \left(\vec{x} + R \hat i\right) \right>
    \label{eq:two_pts}
\end{equation}
and subtract the plaquette in the $\mu$-$\nu$ plane. This defines the \textit{profile} of the flux tube of length $R$ as a function of the transverse displacement $y$:
\begin{equation}
    \rho (R, y) = \frac{F_{01}(R, y)}{G(R)} - \left< \Pi_{01} \right>.
    \label{eq:profile}
\end{equation}
The component $01$, corresponding to the chromo-electric field in the longitudinal direction, is characterized by the strongest signal~\cite{Bonati:2020orj}.

We repeated our measures at different temperature and at different lattice spacings, as listed in Tab.~\ref{tab:num_summary}. For temperatures $T < 0.5 \, T_c$ we adopted the L\"uscher-Weisz ``multilevel'' algorithm to mitigate the signal-to-noise problem.

\begin{table}[]
    \centering
    \begin{tabular}{|c|c|c|c|c|c|}
\hline
$\beta$   & $a \, \sqrt{\sigma}$ & $N_t$ & $T / T_c$ & $N_s$ & $R / a$           \\ 
\hline
8.768     & 0.16702(28)          & 24    & 0.23      & 80    &  9 - 15 \\ \hline
\multirow{6}{*}{10.865} & \multirow{6}{*}{0.13137(86)}
                                 & 60    & 0.11      & 60    &  9               \\
          &                      & 30    & 0.23      & 96    &  9 - 15          \\
          &                      & 20    & 0.34      & 96    &  9, 11, 15, 19   \\
          &                      & 14    & 0.49      & 60    &  7 - 21          \\
          &                      & 10    & 0.68      & 96    & 11 - 21          \\
          &                      &  8    & 0.85      & 96    &  9 - 21          \\ \hline
11.914    & 0.11881(81)          & 11    & 0.68      & 96    & 11 - 21          \\ \hline
\multirow{2}{*}{12.9625} & \multirow{2}{*}{0.10870(62)}
                                 & 36    & 0.23      & 120   & 11 - 17          \\
          &                      & 12    & 0.68      & 96    & 11 - 17          \\ \hline
13.424    & 0.10486(51)          & 10    & 0.85      & 120   &  9 - 21          \\ \hline
14.011    & 0.10046(35)          & 13    & 0.68      & 120   &  9 - 21          \\ \hline
    \end{tabular}
    \caption{Summary of our simulations. In the last column we reported the values of $R$ at which we measured the profile. When we write $R_{min}$ - $R_{max}$ we mean each odd value of $R$ between them, including both extrema.}
    \label{tab:num_summary}
\end{table}

\section{Low Temperature Results}

We attempted different fits to our data for the profile at low temperature, $T \le 0.34 \, T_c$. The model that fits our data with the best accuracy is the Clem model:
\begin{equation}
\label{eq:clem}
    \rho(y) = A^{\rm (Clem)} \, K_0 \! \left( \frac{\sqrt{y^2 + \xi^2}}{\lambda} \right).
\end{equation}
Besides an amplitude $A^{\rm (Clem)}$, the free parameters of the model are two lengths $\lambda$ and $\xi$. This formula was originally introduced to describe the electric field in a type II superconductor~\cite{Clem:1975ohd}, where $\lambda$ is the London length and $\xi$ a variational radius of the Abrikosov vortices. The Clem formula was introduced in the context of flux tube in gauge theories to test the dual superconductor picture of confinement and has been successfully used to fit lattice results for the profile of the flux tube in the (3+1)-dimensional case~\cite{Cea:2012qw, Cea:2017ocq, Baker:2018mhw}.

\begin{figure}
    \centering
    \includegraphics[width=\textwidth]{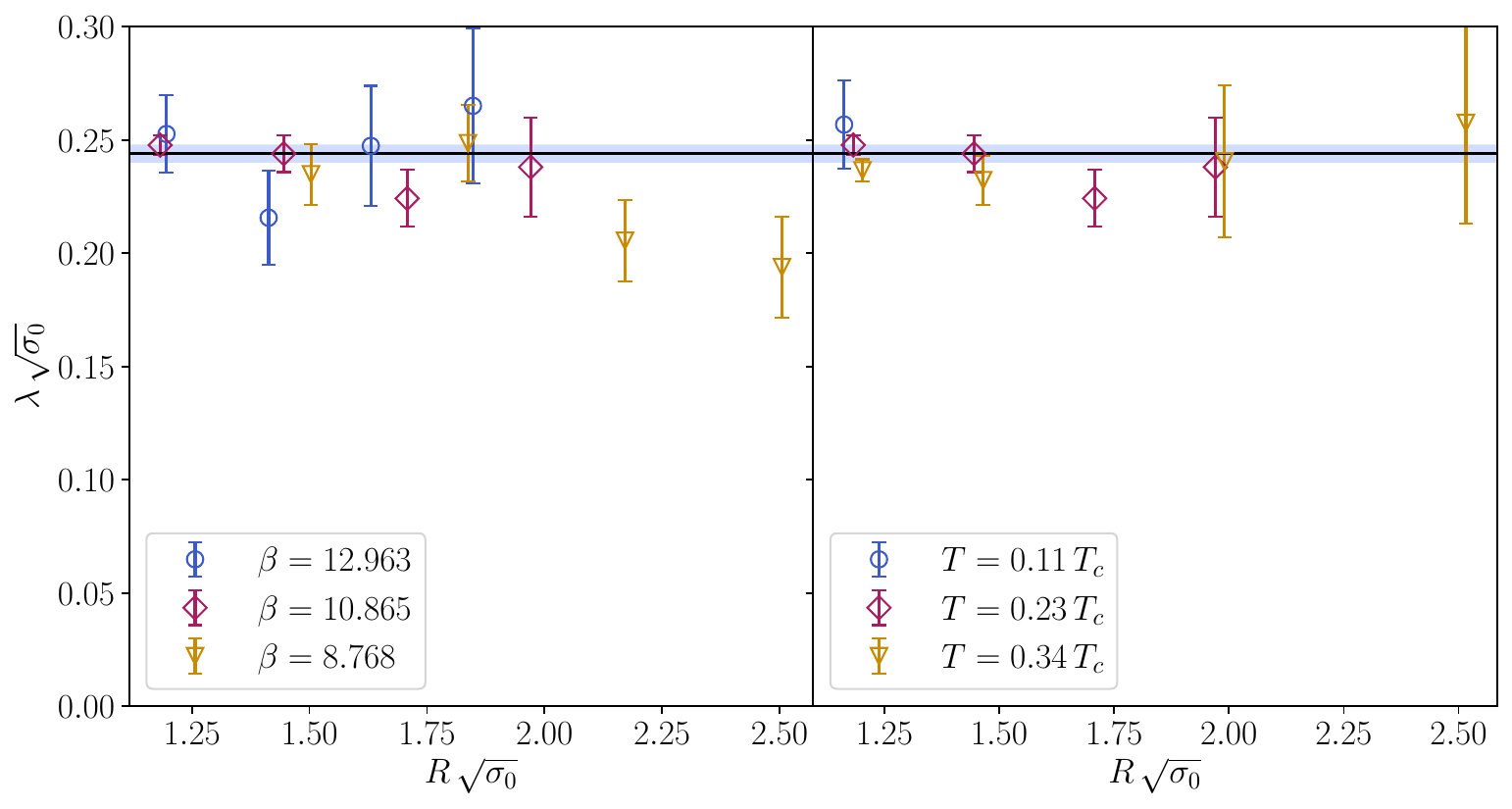}
    \caption{Values of $\lambda$ obtained from the fit using the Clem formula as a model. In both panels the horizontal axis is the length of the flux tube. In the left panel we plotted results from different lattice spacings keeping the temperature constant at $T = 0.23 \, T_c$; in the right one we fixed $\beta = 10.865$ and varied the temperature. The solid line and the shaded band represent our final results and its uncertainty.}
    \label{fig:lambda_consist}
\end{figure}

Our results suggest that $\lambda$ is constant for different flux tube lengths $R$. Repeating the fits a different temperatures, we find compatible results for $\lambda$ and, in units of the string tension $\sigma$, we do not observe any discretization effect, as shown in Fig.~\ref{fig:lambda_consist}. We quote $\lambda \sqrt{\sigma} = 0.244(4)$ as the average result for the value of the intrinsic width from this model.

Our result is similar to the inverse mass of the lightest glueball $M_0$: extracting its value from Ref.~\cite{Athenodorou:2016ebg} we can compute $M_0 \, \lambda = 1.16(3)$, which is, however still significantly different from one. Our value is, instead in agreement with the one from Ref.~\cite{Amorosso:2026mdo}, computed with a completely different approach.

\begin{figure}
    \centering
    \includegraphics[width=0.58333\textwidth]{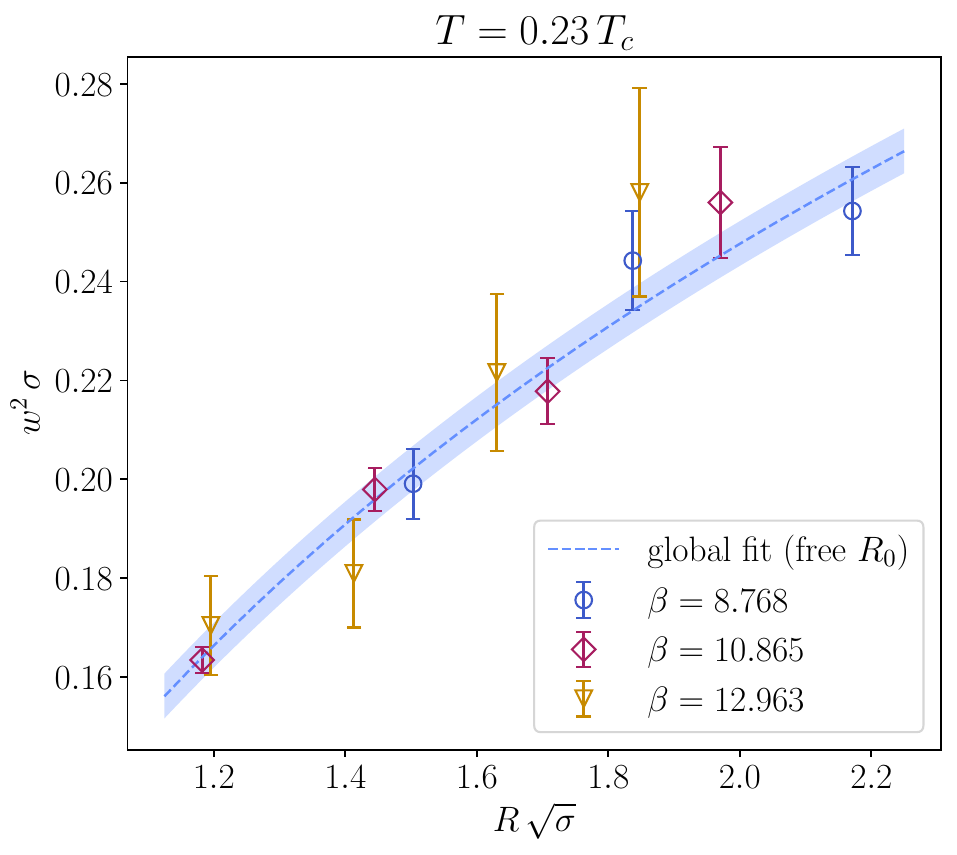}
    \caption{The effective width as a function of the length of the flux tube. The dashed line is the logarithmic broadening expected from effective string theory predictions. The coefficient was fixed to the predicted value, while the vertical offset was fitted to the data.}
    \label{fig:log_broad}
\end{figure}

The parameter $\xi$ clearly exhibits an increasing trend with $R$ instead. This is well in agreement with an effective string theory approach, as it predicts a logarithmic broadening of the flux tube. In particular we can numerically compute the effective square width $w^2$ of the profile, again assuming the Clem formula, and plot the results for the fitted values of $\lambda$ and $\xi$. As we can see in Fig.~\ref{fig:log_broad}, the logarithmic increase is perfectly reproduced.

The dependence of $\xi$ on $R$, however, questions the validity of the dual superconductor picture. Namely, it seems impossible to determine univocally a value of the Ginzburg-Landau parameter $\kappa$, that can be written in terms of the Clem parameters:
\begin{equation}
    \label{eq:clem_ginsland}
    \kappa = \frac{\sqrt{2} \, \lambda}{\xi} \, \sqrt{1 - {K_0}^2 \! \left( \frac{\xi}{\lambda} \right) \Big/ {K_1}^2 \! \left( \frac{\xi}{\lambda} \right)}.
\end{equation}

Assuming a logarithmic broadening of the flux tube (at fixed $\lambda$), one has that $\xi$ goes to infinity in the large $R$ limit and consequently $\kappa$ vanishes. On the other hand, extrapolating our data, $\kappa$ could cross the critical threshold $1 / \sqrt{2}$ for a value of $R$ still in the regime of validity of the effective string description.

\section{High Temperature Results}

However, we can formulate a model for the profile at high temperature, thanks to the Svetitsky-Yaffe mapping. Since the deconfinement phase transition for the $\SU(2)$ model in $(2+1)$ dimensions is in the same universality class as the one of the 2 dimensional Ising model, we can compute correlation functions of Polyakov loops and plaquettes as correlation functions of spin and energy fields in the Ising model.

The relevant correlation functions for our model where computed in Refs.~\cite{Caselle:2006wr, Caselle:2012rp}. From the ratio of correlation functions defined in Eq.~\eqref{eq:profile}, we can write a model for the profile:
\begin{equation}
    \rho(d, y) = A^{\rm(SY)} \, \frac{2 \pi R}{4 l^2} \, \frac{\exp(-l / \lambda)}{K_0\{R / (2 \, \lambda) \}}.
    \label{eq:SY_profile}
\end{equation}
In contrast with the Clem formula, this model presents, beside the amplitude $A^{\rm(SY)}$, only one free length scale (the intrinsic width $\lambda$), since the role of the second length scale is played by the distance $R$ between the Polyakov loops. Also note that the dependence of the amplitude on the length of the flux tube has been factorized, so that $A^{\rm(SY)}$ should not depend on $R$. Moreover, the mapping predicts that the length scale in the three-point function should be identical to the two-point one. Thus, we compare the value of $\lambda$ fitted from the profile to the length we extract from fitting the Polyakov loop correlator (see Refs.~\cite{Caselle:2021eir, Caselle:2024zoh}), or equivalently check that
\begin{equation}
    \lambda = \frac{1}{2 \, E_0},
    \label{eq:intrwidth_ground}
\end{equation}
where $E_0$ is the ground state in the effective string theory.

We fitted out data of the profile with the model based on the Svetitsky-Yaffe mapping finding good agreement for $T \ge 0.68 \, T_c$. Since neither of the parameters in the model depends on the length of the flux tube, we are able to describe the data obtained on a given lattice with a single combined fit. Finally we compare the values of $\lambda$ we extract from the profile to the ground state we obtain from the correlator of Polyakov loops, verifying Eq.~\eqref{eq:intrwidth_ground} within at most three standard deviations.

Also the linear broadening of the flux tube, predicted by the effective string theory, can be probed within the Svetitsky-Yaffe mapping. To do so, we compute the corresponding effective width and expand it in series of $\lambda / R$:
\begin{equation}
    \label{eq:SY_linearbroad}
    w^2(R) = \frac{\lambda R}{2} - \frac{\lambda^2}{2} + \mathcal{O}\left( \frac{\lambda^3}{R} \right) .
\end{equation}

\begin{figure}
    \centering
    \includegraphics[width=0.583333\textwidth]{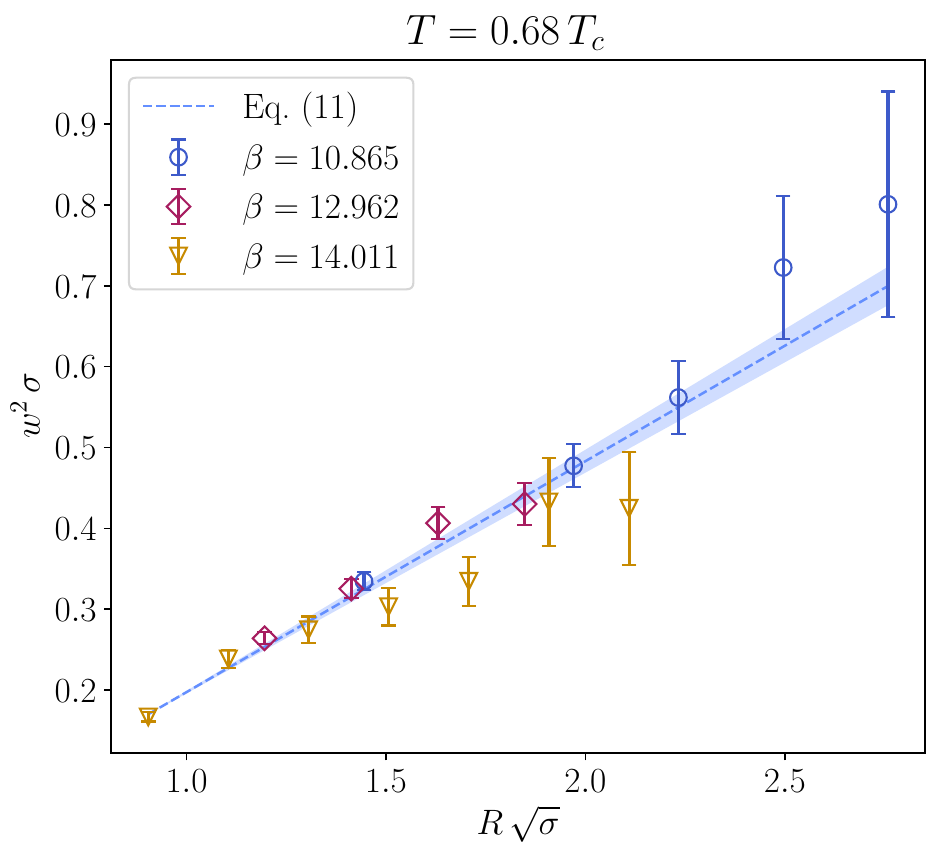}
    \caption{Square of the total width from the fit assuming the SY mapping of our data at $T = 0.68 \, T_c$. The dashed line with the confidence band is the expansion from Eq.~\eqref{eq:SY_linearbroad}. The input value of $\lambda$ is the one extracted from the Polyakov loop correlator $\lambda = 1 / (2 E_0)$.}
    \label{fig:compare_binderwidt_07tc}
\end{figure}

When we substitute $\lambda = 1 / (2 E_0) = T / (2 \sigma(T))$, the formula perfectly matches the linear broadening formula from Ref.~\cite{Allais:2009uos}:
\begin{equation}
    w^2(R) = \frac{T \, R}{4 \, \sigma(T)} + \dots \, .
\end{equation}
The agreement can be appreciated in Fig.~\ref{fig:compare_binderwidt_07tc}.

\begin{figure}
    \centering
    \includegraphics[width=0.583333\textwidth]{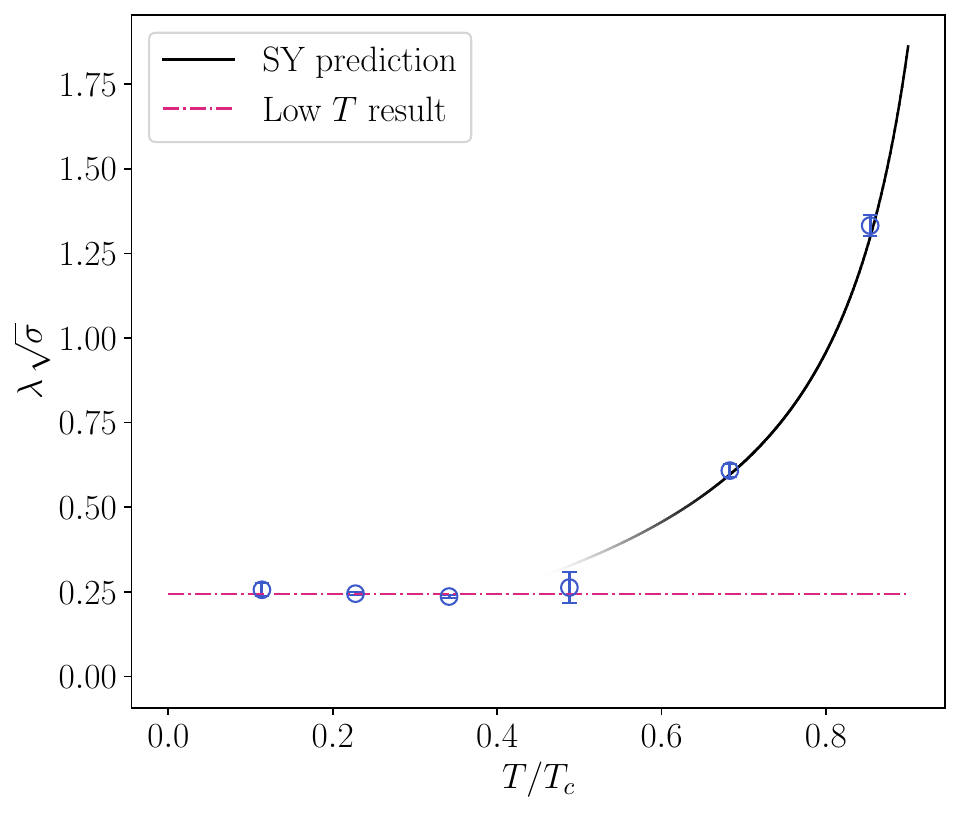}
    \caption{The values of $\lambda$ as a function of the temperature. The red dash-dotted line is our estimation at low temperature. The black solid line is the SY prediction, using the fit from Ref.~\cite{Caselle:2024zoh} for $E_0$.}
    \label{fig:lamb_vs_temp}
\end{figure}

Neither the Svetitsky-Yaffe nor the Clem formula provide a good fit to our data at $T = 0.49 \, T_c$: here a fit of the tails of the profile with an exponential gives our the best estimation of $\lambda$. Finally, we can plot the intrinsic width of the flux tube as a function of the temperature, in units of the string tension. The result is shown in Fig.~\ref{fig:lamb_vs_temp}.

Two regimes can clearly be identified: at low temperature the intrinsic width is constant and related to the mass of some excitation in the bulk. Approaching the deconfinement temperature, instead, the intrinsic width diverges as the string ground state tends to vanish.

\section{Conclusions}
In this contribution we presented our updated results on the intrinsic width of the flux tube in (2+1)-dimensional Yang-Mills theory with $\SU(2)$ gauge group. We succeed to identify the intrinsic width as the characteristic length of the exponentially decaying tails of the flux tube profile, which are the most evident deviations from a simple Gaussian shape. Such a length does not show a dependence on the length of the flux tube and, for a subset of the values of the temperature we investigated, we checked that it has no discernible discretization effects. We find that the intrinsic width of the flux tube has a constant value at low temperature, while it grows approaching the deconfinement phase transition.

At low temperature, our data can be fitted by the Clem model which originated form the dual superconductor picture; however, such a picture does not seem to be completely consistent, since it a meaningful value of the Ginzburg-Landau parameter cannot be extracted. Using this approach we compute instead our estimation of the intrinsic width at low temperature, $\lambda \sqrt{\sigma} = 0.244(4)$. Notably, this estimation is similar (but not compatible) with the inverse of the mass of the lightest glueball. Furthermore, it is in very good agreement with the length scale obtained studying the entanglement entropy of the flux tube~\cite{Amorosso:2026mdo}.

At high temperature, instead, we test the predictions from the Svetitsky-Yaffe mapping for the profile of the flux tube and find very good agreement already at temperature as low as $T = 0.68 \, T_c$. The mapping predicts the value of the intrinsic width to be related to the ground state energy in the effective string theory. We check that this prediction is verified within at most three standard deviations from our data. Also the expected linear broadening of the flux tube in this regime is successfully tested.

\section*{Acknowledgements}
We thank O.~Aharony, R.~Amorosso, C.~ Bonati, T.~Canneti, M.~Panero, M.~Pepe for useful discussion. All the authors acknowledge support from the SFT scientific initative of INFN, The work of M.~C., A.~N., D.~P.~and L.~V.~is supported by Simons Foundation grant 994300 (Simons Collaboration on Confinement and QCD Strings). A.~N.~acknowledges support by the European Union - Next Generation EU, Mission 4 Component 1, CUPD53D23002970006, under the Italian PRIN “Progetti di Ricerca di Rilevante Interesse Nazionale – Bando 2022” prot. 2022ZTPK4E. Numeric computations were performed on the ``Leonardo'' machine at CINECA, based on the agreement with INFN, under the project INF25\_sft.

\bibliographystyle{JHEP}
\bibliography{biblio}

\end{document}